\documentclass[floatfix,lengthcheck,showpacs,amssymb,amsmath,amsfonts,twocolumn]{aastex63}
\usepackage{CJK}
\usepackage{lineno}
\usepackage[utf8]{inputenc}
\usepackage{color}
\usepackage{graphicx}
\usepackage{float}
\usepackage{subfigure}
\usepackage{physics}
\usepackage{amsmath}
\usepackage{afterpage}
\usepackage{acronym}
\usepackage{multirow}
\usepackage{tikz}
\usetikzlibrary{arrows,shapes,trees,decorations.pathreplacing}
\usepackage{import}
\usepackage{svn-multi}
\usepackage{lineno}
\usepackage{hyperref}
\usepackage{gensymb}
\hypersetup{
    colorlinks=true,
    linkcolor=blue,
    filecolor=magenta,      
    urlcolor=cyan,
}

\newcommand{\ICTS}{\affiliation{International Centre for Theoretical Sciences, Tata Institute of Fundamental Research, Bangalore 560089, India}}
\newcommand{\AEI}{\affiliation{Max-Planck-Institut f{\"u}r Gravitationsphysik (Albert-Einstein-Institut), D-30167 Hannover, Germany}}
\newcommand{\UniHannover}{\affiliation{Leibniz Universit{\"a}t Hannover, D-30167 Hannover, Germany}}

\begin{document}
\begin{CJK*}{UTF8}{gbsn}
\title[]{Detecting Baryon Acoustic Oscillations with third generation gravitational wave observatories}

\correspondingauthor{Sumit Kumar}
\email{sumit.kumar@aei.mpg.de}
\author[0000-0002-6404-0517]{Sumit Kumar}
\AEI{}
\UniHannover{}
\author[0000-0002-4103-0666]{Aditya Vijaykumar}
\ICTS{}
\author[0000-0002-1850-4587]{Alexander H. Nitz}
\AEI{}
\UniHannover{}


\begin{abstract}
We explore the possibility of detecting Baryon Acoustic Oscillations (BAO) solely from gravitational wave observations of binary neutron star mergers with third generation (3G) gravitational wave (GW) detectors like Cosmic Explorer and the Einstein Telescope. These measurements would provide a new independent probe of cosmology. The detection of the BAO peak with current generation GW detectors (solely from GW observations) is not possible because i) unlike galaxies, the GW mergers are poorly localized and ii) there are not enough merger events to probe the BAO length scale. With the 3G GW detector network, it is possible to observe $\sim \mathcal{O}(1000)$ binary neutron star mergers per year localized well within one square degree in the sky for redshift $z \leq 0.3$. We show that 3G observatories will enable precision measurements of the BAO feature in the large-scale two-point correlation function; the effect of BAO can be independently detected at different reshifts, with a log-evidence ratio of $\sim$ 23, 17, or 3 favouring a model with a BAO peak at redshift of 0.2, 0.25, or 0.3, respectively,  using a redshift bin corresponding to a shell of thickness $150 h^{-1}$ Mpc.
\end{abstract}

\section{Introduction}
The catalog of gravitational wave (GW) transients from compact binary mergers has grown considerably \citep{LIGOScientific:2018mvr, LIGOScientific:2020ibl, Nitz:2019hdf, Nitz:2021uxj, Venumadhav:2019lyq} since the first detection of gravitational waves  from the merger of the binary black hole GW150914 \citep{Abbott_2016}. This growing catalog of mergers has already revolutionized our understanding of the astrophysical rates and populations of compact objects, and has enabled precision tests of general relativity and cosmology \citep{LIGOScientific:2020tif, LIGOScientific:2019zcs}. The sensitivity of the current ground-based GW detector network to compact binary mergers is expected to improve when the LIGO \citep{LIGOScientific:2014pky}, Virgo \citep{VIRGO:2014yos} and KAGRA \citep{KAGRA:2020tym} detectors undergo upgrades \citep{KAGRA:2013rdx}, and also with the construction of new detectors like LIGO-India \citep{Saleem:2021iwi}. Additionally, third generation (3G) detectors such as Einstein Telescope (ET) \citep{Sathyaprakash:2012jk} and Cosmic Explorer (CE) \citep{Reitze:2019iox} will have an order-of-magnitude better strain sensitivity and will also be able to probe lower GW frequencies. It is also expected that they will localize most mergers within a few square degrees, while detecting hundreds of thousands of binary mergers each year \citep{Mills:2017urp}. A number of precision tests of astrophysics and cosmology will be enabled as a result---for instance, studying the spatial distribution of a large number of well-localized sources, one can probe the large scale distribution of matter in the universe \citep{Vijaykumar:2020pzn,Mukherjee:2020hyn, Libanore:2020fim, Mukherjee:2020mha, 2021Herrera}. These probes using GW observations could confirm if the distribution of GW mergers indeed track the galaxy distribution, and can provide an independent probe to the features mostly attributed to galaxy or quasar population e.g. clustering bias \citep{Kaiser:1984sw}.

In this study, we investigate the possibility of probing another feature of the cosmological large scale structure---baryon acoustic oscillations---with third generation GW detectors. The detection of BAO peak with GW events can open up a complimentary window to probe cosmological parameters. GW detector networks are sensitive to mergers happening in all directions in the sky. For redshift $z<0.3$, where we expect localization of large number of GW mergers to be precise enough (within one square degree), all the observed GW mergers can be used to probe BAO feature and, unlike galaxy surveys, we won't be limited by the survey volume.

The layout of the paper is as follows: in Section \ref{sec:cosmo} we give a brief overview of cosmological probes with GW observations, and motivate baryon acoustic oscillations as an independent probe of large scale structure.  In Section \ref{sec:third-gen}, we describe the configurations of the 3G GW detector network used in this study. We describe our methodology to generate mock binary neutron star observations in Section \ref{sec:methods-and-results}, along with estimates of the measurability of the BAO feature in the correlation function. We end by summarizing our results and future directions in Section \ref{sec:summary}.

\section{Cosmology and gravitational waves}\label{sec:cosmo}
Data from various cosmological surveys indicate that the evolution and current state of the Universe is best described by the standard model of cosmology, also referred to as the $\Lambda$CDM model \citep{SupernovaSearchTeam:1998fmf}. This model includes dark energy (described by the cosmological constant $\Lambda$ in Einstein's equations) as the dominant component, along with dark matter (a pressure-less fluid which interacts with standard model particles purely through gravitational forces), and baryonic matter (which includes directly observable matter such as galaxies and the intergalactic medium). Given a cosmological model and a set of parameters, one can derive the relation between the distance to an astronomical object, and the cosmological redshift $z$ due to cosmic expansion.  Conversely, independent measurements of the distances and $z$ from observations can be turned into into measurements of the cosmological model parameters.

In the last few years, a 4.4$\sigma$ discrepancy has been reported between the value of the Hubble parameter $H_0$ measured using early universe \citep{Planck:2018vyg} and late universe \citep{Riess:2019cxk} probes, hinting either at unknown systematics in the measurements, or at a ``Hubble Tension'' and possible deviation from the $\Lambda$CDM paradigm. The independent measurement of the Hubble parameter using GWs from compact binaries is ideally suited to provide more clarity in this regard. The characteristic luminosity of GW sources provides a direct measurement of the luminosity distance out to the sources \citep{Schutz:1986gp}. If the redshift of these sources can be measured using any other methods like the detection of an electromagnetic counterpart \citep{Holz:2005df, Dalal:2006qt, Nissanke:2013fka} statistical identification of the host galaxy using a galaxy catalog \citep{DelPozzo:2011vcw,Chen:2017rfc}, a measurement of the tidal parameter \citep{Messenger:2011gi,Chatterjee:2021xrm}, or a physical scale in the mass distribution of sources \citep{Farr:2019twy, Ezquiaga:2020tns, You:2020wju}, one can make a measurement of the Hubble parameter. It is expected that a measurement accuracy of $\sim 4.4\%$ can be reached with $\sim 250$ binary neutron star merger detections \citep{Gray:2019ksv}.

Another avenue of study in cosmology where GW observations show promise is their use as tracers to study the large-scale structure of the Universe. Similar to how galaxy surveys are used to probe large scale clustering, a population of GW sources can be used to probe the cosmological large scale structure by either the three-dimensional autocorrelation of the sources \citep{Vijaykumar:2020pzn}, or by cross-correlating the sources  with other tracers of large-scale structure \citep{Bera:2020jhx, Mukherjee:2020hyn, Libanore:2020fim, Mukherjee:2020mha}. These allow for constraints to be put on the large-scale bias of gravitational wave events $b_\mathrm{GW}$, as well as the parameters of the standard $\Lambda$CDM model of cosmology. 

\begin{figure}[t]
  \centering
    \includegraphics[width=\columnwidth]{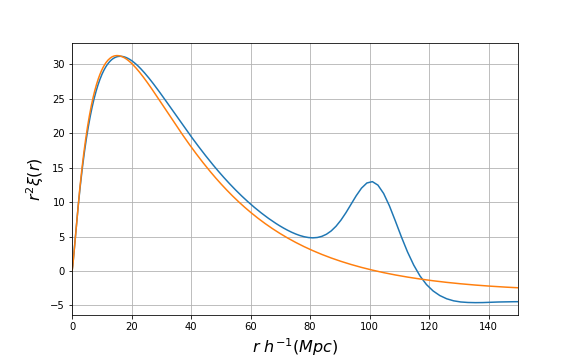}
\caption{An example of two-point correlation function $\xi(r)$ as a function of comoving distance $r$. Here we show two correlation functions: i) $\xi(r)$ showing a BAO feature at the scale of $\sim$ 100 h$^{-1}$Mpc is calculated using transfer function prescribed by Eisenstein and Hu \citep{Eisenstein_1998}, and ii) $\xi(r)$ without the BAO feature is calculated using the BBKS \citep{1986ApJ...304...15B} transfer function. We assume $\Lambda$CDM cosmological model parameters consistent with the Planck 2015 data \citep{Planck:2015fie}. We multiply the two point correlation function $\xi(r)$ with $r^2$ on the vertical axis for a better visualization of the BAO peak. The units on the horizontal axis  are $h^{-1}$ Mpc where $h$ is defined in terms of the Hubble constant $H_0 = 100h$ km s$^{-1}$ Mpc$^{-1}$}
\label{fig:corrfunc}
\end{figure}

\begin{table*}
    \caption{The specifications of each detector (location, noise curves, low frequency cutoff $f_\mathrm{low}$) considered in this study. For CE detector, subscript (1, 2) represents (early, late) noise sensitivity curves and superscript (U, A) represents location of these detector (USA, Australia). These detectors configuration for CE and ET are taken from \citep{Nitz_2021}
    }
    \label{table:detectors}
\begin{center}
\begin{tabular}{llllcc}
Abbreviation & Observatory  & $f_{\textrm{low}}$ &  Noise Curve & Latitude & Longitude \\ \hline
$C_1^U$ & Cosmic Explorer USA & 5.2 & CE1 & 40.8 & -113.8   \\
$C_1^A$ & Cosmic Explorer Australia & 5.2 & CE1 & -31.5 & 118.0  \\
$C_2^U$ & Cosmic Explorer USA & 5.2 & CE2 & 40.8 & -113.8    \\
$C_2^A$ & Cosmic Explorer Australia & 5.2 & CE2 & -31.5 & 118.0   \\
$E$ & Einstein Telescope & 2 & ET-D Design & 43.6 & 10.5  \\

\end{tabular}
\end{center}
\end{table*}

In this work, we ascertain the possibility of probing another feature in large-scale clustering of matter, namely baryon acoustic oscillations (BAO) \citep{Sakharov:1966aja, Peebles:1970ag, Sunyaev:1970eu, Eisenstein:1997ik}. BAO are imprints left by early-time sound waves in the Universe on the late-time distribution of matter. In the early Universe (at redshifts $ > 1089$), high temperatures prevented the existence of bound atoms, and the primordial gas existed as ionized plasma. Free electrons in this plasma interacted with photons via Thomson scattering, thus coupling the baryons, electrons and photons into an effective fluid. The competing forces of electromagnetic radiation pressure and gravity in this fluid generated perturbations, thus setting up sound waves in the fluid.  During the epoch of recombination ($z\sim1089$), the Universe cooled down enough for stable atoms to form---this thwarted the Thomson scattering, and destroyed the coupling. The photons then free-streamed and formed what we now know as the Cosmic Microwave Background (CMB), while the perturbations froze at a certain scale. As the Universe evolved and formed structures, this scale got imprinted on the distribution of halos and galaxies in the Universe at late times, appearing as a peak in the two point correlation function. For reviews on BAO, see \citep{Bassett:2009mm, Weinberg:2013agg}.

The first confident signature of BAO from galaxy surveys came from the $3.4 \sigma$ detection in the large-scale correlation function of luminous red galaxies (LRG) from Sloan Digital Sky Survey (SDSS) Data Release 3 \citep{SDSS:2005xqv}. These measurements have been confirmed by other samples like the 6-degree Field Galaxy Survey \citep{Beutler:2011hx}, the WiggleZ Dark Energy Survey \citep{Blake:2011en, Blake:2011wn}, and most recently by the SDSS-IV extended Baryon Oscillation Spectroscopic Survey (eBOSS) \citep{eBOSS:2020yzd,Bautista:2020ahg}.

The BAO signature can be seen in the correlation function as a peak at a comoving scale of $\sim$100 $h^{-1}$ Mpc. In figure \ref{fig:corrfunc}, we show the three-dimensional correlation function $\xi(r)$ calculated using a transfer function fit provided by Eisenstein-Hu \citep{Eisenstein_1998} (with BAO feature) and by Bardeen et al \citep{1986ApJ...304...15B} BBKS (without BAO feature). This signature can also be captured by the two-point angular correlation function (2PACF) $w(\theta)$.  Given a galaxy survey, one can estimate the correlation function using various estimators, most notably the Landy-Szalay estimator \citep{Landy:1993yu}. 
For the localization volumes of typical binary mergers, the errors along the radial direction are larger compared to errors in angular direction when those errors are converted into comoving length scales. It is hence convenient to measure the 2PACF from GW merger events, provided that the radial uncertainties are not large enough to smear away information in the correlation function at the scales of interest. In general, one needs to take into account the smearing of the measured correlation function due to localization errors \citep{Vijaykumar:2020pzn}, and projection effects \citep{Limber:1954zz} to track the effective shape of 2PACF $w(\theta)$.

\begin{figure}[h]
  \centering
    \includegraphics[width=\columnwidth]{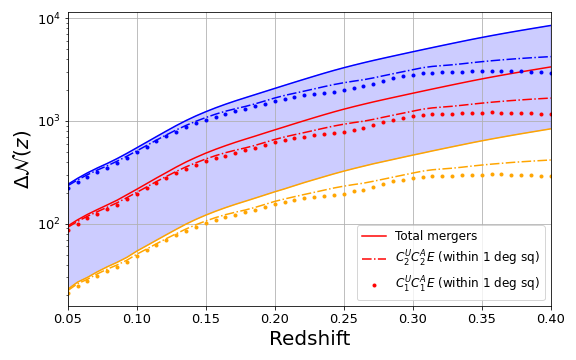}
\caption{The number of BNS mergers per year in a shell of thickness 150 h$^{-1}$ Mpc as a function of redshift. Solid lines represents total number of mergers, dotted-dashed and dashed lines represent the BNS mergers with sky localization within 1 square degree in that shell for two detector network (See text for explanation). { The red lines represent the mean number of BNS mergers in the shell per year, corresponding to the mean value of merger rate. Similarly, the blue (yellow) lines represent the upper (lower) limits on the number of BNS mergers corresponding to the upper (lower) limits of BNS merger rate. The shaded blue region represents the range of values that number of mergers can take between these upper and lower limits. Solid lines represent total number of mergers, while dashed-dotted and dotted lines represent the total number of detectable events with sky localization errors $< 1 ~\mathrm{deg}^2$ for different detector networks considered.}
}
\label{fig:radec_distribution}
\end{figure}

\begin{figure*}[htb!]
  \centering
    \hspace*{-.5cm}\includegraphics[width=2.5\columnwidth]{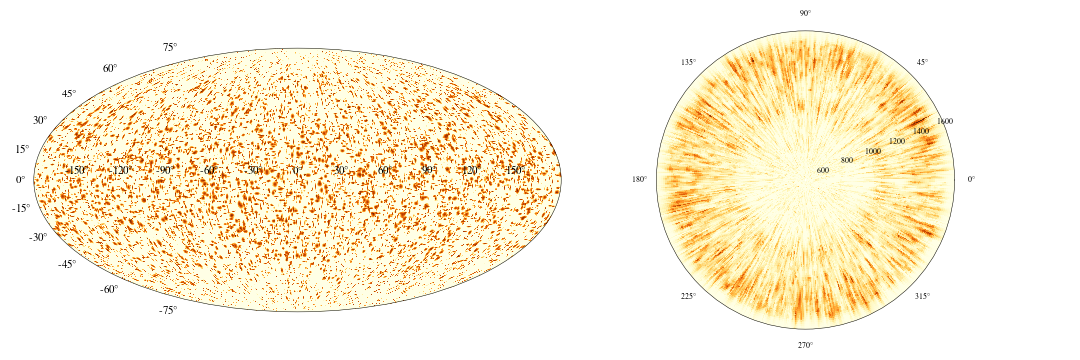}
\caption{A realization of combined posterior field for the marginalized localization posterior from the simulation of of BNS events using 3G detector network. Left panel shows the marginalized posteriors for right ascension (RA) and declination (dec) angles. Right panel shows the marginalized posteriors on RA and comoving distance (along radial direction).}
\label{fig:bns_population}
\end{figure*}

\section{Third generation detector network}\label{sec:third-gen}
The proposed third generation (3G) detectors such as CE~\citep{cewhitepaper,2021arXiv210909882E} and ET~\citep{et} are expected to be operational sometime in next decade (2030s). CE is proposed to be built in two stages with upgrade consists of increasing design complexity and better sensitivity, known as CE1 and CE2 ~\citep{Hall:2020dps}. ET is proposed to have good sensitivity at low frequency ~\citep{Hild:2010id}; we consider the design sensitivity $f_\mathrm{low}$ = 2Hz of ET for this study. The location of these detectors are not yet finalized but we use a fiducial location for these detectors: one CE in USA and the other CE in Australia which provides a long baseline. These fiducial detector locations have also been used in previous works \citep{2019CQGra..36v5002H, Nitz_2021} We consider the location of ET to be in Europe. Table \ref{table:detectors} lists the properties of the detectors we consider in this study. 
In this study, we focus on the localization capabilities of the 3G detector network only. Any 2G detector(s) added to the network would only further enhance the localization capabilities of the network.
We consider following detector network configurations: \\
i) {$C_1^UC_1^AE$}: Two CE detectors (One in the USA and the other in Australia) and ET (in Europe), where the CE detectors have the early phase design sensitivity CE1, and \\
ii) {$C_2^UC_2^AE$}: Same as above, but a CE with second phase design sensitivity CE2.

Although we examine these specific configurations of the worldwide detector network, we do not expect other detector network configurations to change the distribution of localization errors of BNS events significantly as long as they include several next-generation observatories.

\section{Simulations and Results}\label{sec:methods-and-results}
Next generation detectors will significantly improve the localization for both BNS and BBH mergers, and due to the higher intrinsic merger rates, we expect to get a much larger number of BNS events with highly precise localization volumes at low redshift ($z < 0.3$). Although we only consider BNS simulations in this study, our method can be readily generalized to BBHs. 

We create a fiducial universe containing localization posteriors for BNS events observed with 3G detector networks, and we call it a ``BNS catalog''. To make such a BNS catalog, we create a realization of the universe containing large number of galaxies (fiducial galaxy catalog), a (randomly selected) small fraction of which can act as the host galaxies to BNS events. These galaxies need to be distributed spatially in such a way that underlying correlation function contains the BAO peak as shown in figure \ref{fig:corrfunc}.

\subsection{BNS population distribution}
To generate a realistic population of BNS, we use the Madau-Dickinson star formation rate (SFR) $\psi(z)$ \citep{Madau:2014bja},
\begin{equation}
    \psi(z) = 0.015\frac{(1+z)^{2.7}}{1+[(1+z)/2.9]^{5.6}} ~\rm{M_\odot yr^{-1} Mpc^{-3}} \label{eqn:SFR}
\end{equation}
\noindent
We assume that the local formation rate of the BNS is proportional to the SFR. To get the merger rate, the SFR is corrected with a delay time distribution $p(t_D) \sim 1/t_D \sim 1/(t-t_f)$ where $t_f$ is the formation time of the binary.
\begin{equation}
    \Psi(z) = \int_{z_f}^z \psi(z') \ P\qty(t(z')-t_f) \dd{z'} \label{eqn:SFR_TD} 
\end{equation}
\noindent
This choice of the delay time distribution is motivated by classical isolated binary evolution models \citep{OShaughnessy:2009szr, Dominik:2012kk}. We normalize \ref{eqn:SFR_TD} such that $\Psi (z=0)$ gives us the local merger rate of 320 $\mathrm{yr}^{-1} \mathrm{Gpc}^{-3}$, the median estimated merger rate of BNS mergers from GWTC-2 \citep{LIGOScientific:2020kqk}.  In the detector frame, the number density of of BNS mergers $dN/dz$ is related to source frame merger rate $\Psi(z)$ by following relation,
\begin{equation}
    \frac{dN}{dz} = \frac{dV_c}{dz}\frac{\Psi(z)}{1+z} \label{eqn:rate_observer}
\end{equation}

Where $V_c$ is the comoving volume. We integrate \ref{eqn:rate_observer} in a given redshift bin and estimate the total number of BNS mergers $\Delta \mathcal{N}(z)$ expected in that redshift bin from 3G detectors. The results we thus obtain are consistent with \citep{Mills:2017urp}.

\subsection{Parameter estimation}

To estimate the localization posterior for each simulated BNS source, we make use Bayesian parameter estimation using the publicly available code PyCBC Inference~\citep{Biwer:2018osg}. We distribute non-spinnning BNS sources assuming the source frame component masses to be equal to 1.4 $M_\odot$. Since the mass distribution of neutron stars is narrow, we do not expect the results of the study to differ significantly with any other mass distributions for BNS sources.
We assume that the sources are distributed isotropically in sky and orientation for inclination angle, and uniformly in comoving distance. The redshift (or distance) distribution can be obtained by rescaling base population to desired rate as a function of redshift such as in \ref{eqn:rate_observer}. We use heterodyne likelihood model ~\citep{Cornish:2010kf,Finstad:2020sok,Zackay:2018qdy} to estimate the likelihood function. We choose following parameters to vary in parameter estimation: {chirp mass: $\mathcal{M}$, mass ratio: $q$, ($q>1$), inclination angle, luminosity distance: $D_L$, Right Ascension: RA, declination: dec, polarization angle, merger time:$t_c$}. We use uniform priors on $\mathcal{M}$ (detector frame), $q$, and $t_c$ and isotropic priors for RA, dec, inclination angle, and polarization. For distance, we choose a prior uniform in comoving volume. We use the \textsc{TaylorF2} waveform model \citep{Blanchet:1995ez, Faye:2012we} implemented in \textsc{LALSuite} \citep{lalsuite} to simulate our signal in gaussian noise, and for signal recovery while estimating source parameters. \textsc{TaylorF2} excludes the merger from the analysis but we still recover significant signal to noise ratio (SNR) due to long signal length and enhanced low frequency sensitivity of 3G detectors. Due to the significantly low frequency cutoff of ET ($f_\mathrm{low} \sim 2$ Hz), the length of the signal is very long and hence we take earth rotation effects into account.  We sample the signal at 1024 Hz, and introduce a high frequency cut-off of 512 Hz for evaluation of the likelihood function in order to reduce computational costs. Ideally, the high-frequency cutoff should be much larger, but this does not cause a significant loss in SNR compared to the full signal, and we are still able to get highly localized posteriors for the redshift range we are interested in. To sample over the parameters, we use a sampler based on a dynamical nested sampling algorithm \citep{Higson_2018,10.1214/06-BA127} implemented in software package \textsc{dynesty} \citep{speagle:2019}.

\subsection{Methodology}
For the purposes of this study, we assume that BNS events are hosted in galaxies and hence they trace the underlying galaxy distribution. To create a realization of BNS events that trace the galaxy distribution, we first choose a shell centred around the redshift we are interested in, and generate an underlying fiducial galaxy catalog. The density of BNS events selected depends on the total number of mergers expected in the shell, with the additional condition that they should be localized within one square degree. Figure \ref{fig:radec_distribution} shows the number of events that satisfy this criterion as a function of redshift.

To generate the fiducial galaxy catalogs, we use publicly available code {lognormal\_galaxies} \citep{Agrawal_2017}. The input power spectrum is calculated using the Eisenstein and Hu transfer function \citep{Eisenstein:1997ik} which contains the BAO peak. We assume standard $\Lambda$CDM cosmology with parameters consistent with Planck results \citep{Planck:2015fie}. After construction of the galaxy catalog, host galaxies are chosen randomly and localization posteriors are placed according to the errors obtained from simulations. In figure \ref{fig:bns_population} we illustrate a realization of one such BNS catalog with marginalised posteriors for localization parameters. Each BNS catalog consists of $N$ posterior samples combined to give posterior field $\mathcal{P} = \sum_{i=1}^N \mathcal{P}_i(RA, dec, D_C)$, where $\mathcal{P}_i$ is individual localization posterior for RA, dec, and comoving distance: $D_c$. 

Once we have a BNS catalog at given redshift, to extract the BAO peak, we focus on a shell of thickness $\approx$ 150 $h^{-1}$ Mpc at redshifts $z = \{0.2, 0.25, 0.3\}$. We use the following algorithm for extracting BAO peak:
\begin{figure}[t]
  \centering
    \includegraphics[width=\columnwidth]{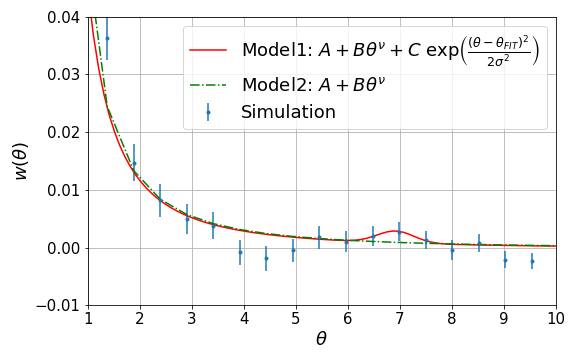}
\caption{2PACF recovery is shown here for a realization at the redshift $z=0.3$. We also show the fit to the data using the model described in the text. Input value for $\theta_{BAO}$ for $z=0.3$ is 6.9 degrees. We estimate the difference in log evidence for both the models $ln\frac{Z1}{Z2} = 2.59$ indicating that the model with a BAO peak is favoured compared to the model without a BAO peak. {The errors are obtained by averaging 1000 catalogs which account for sampling bias due to selection of galaxies for BNS merger events, cosmic variance, and due to scatter in the localization posteriors.}}

\label{fig:w_theta_recovered}
\end{figure}

\begin{figure}[t]
  \centering
    \includegraphics[width=\columnwidth]{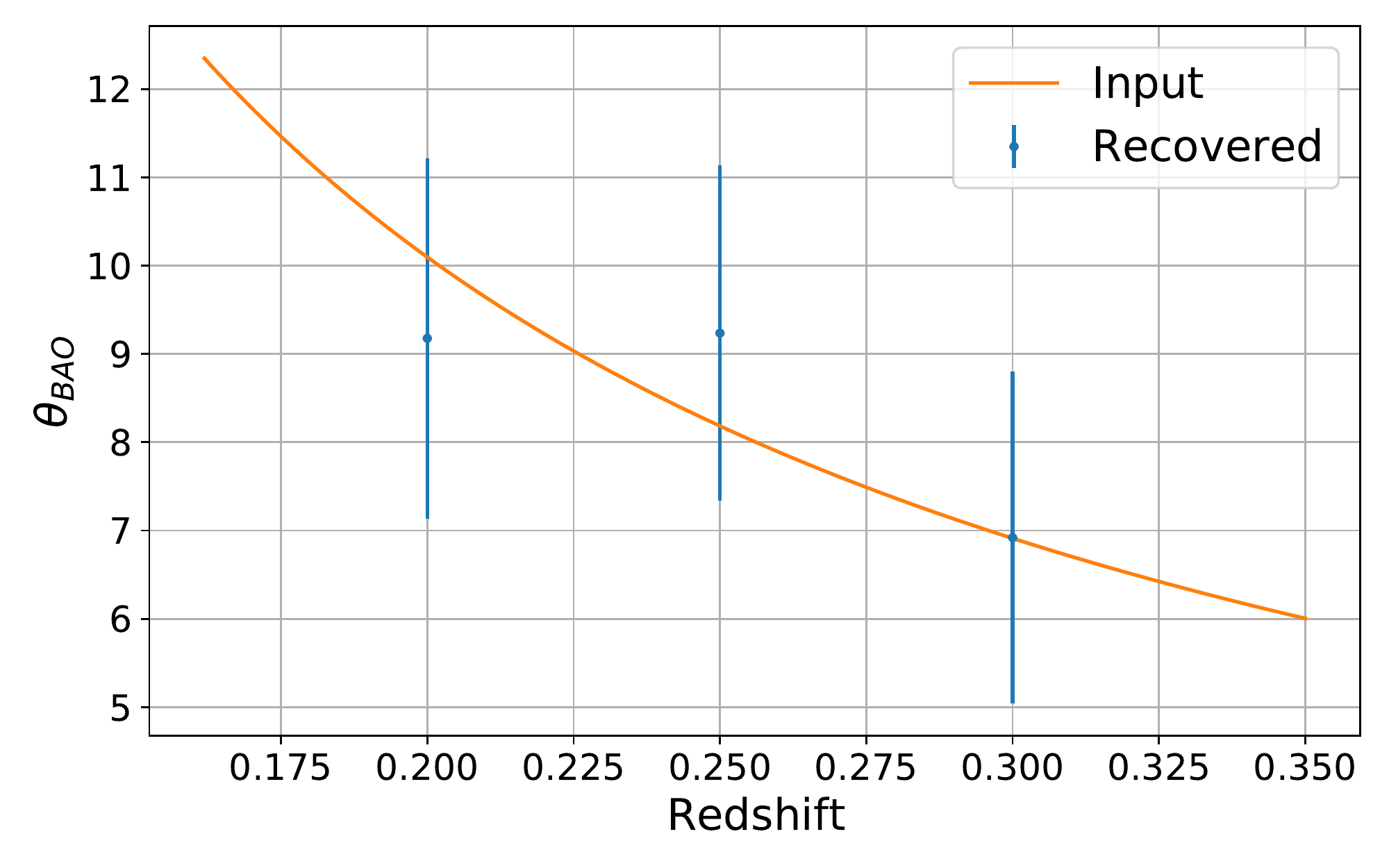}
\caption{The recovery of BAO scale at different redshifts. The solid continuous line shows the angular BAO scale as a function of redshift. The errors on the recovered BAO scale are estimated from averaging over 1000 catalogs to account for cosmic variance as well as statistical errors due to selecting host galaxies for BNS merger events from large galaxy catalogs. We estimate the difference in the log evidence for fit function for two models (with BAO against without BAO). For all the redshifts, the model with BAO peak is favoured (see text).  }
\label{fig:bao_recovery}
\end{figure}

\begin{figure*}[t]
\centering
\setkeys{Gin}{width=0.32\linewidth}
\includegraphics{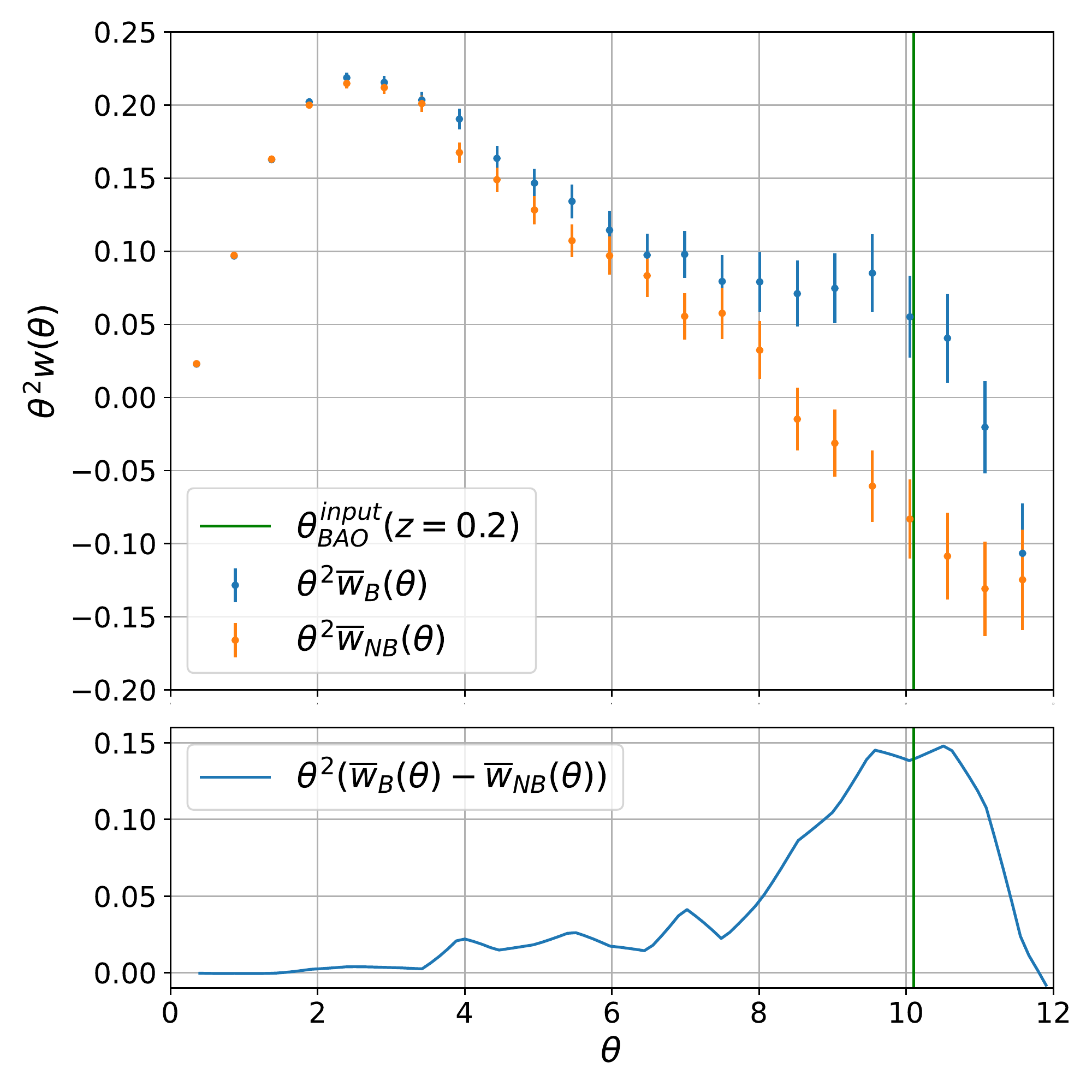}
\hfill
\includegraphics{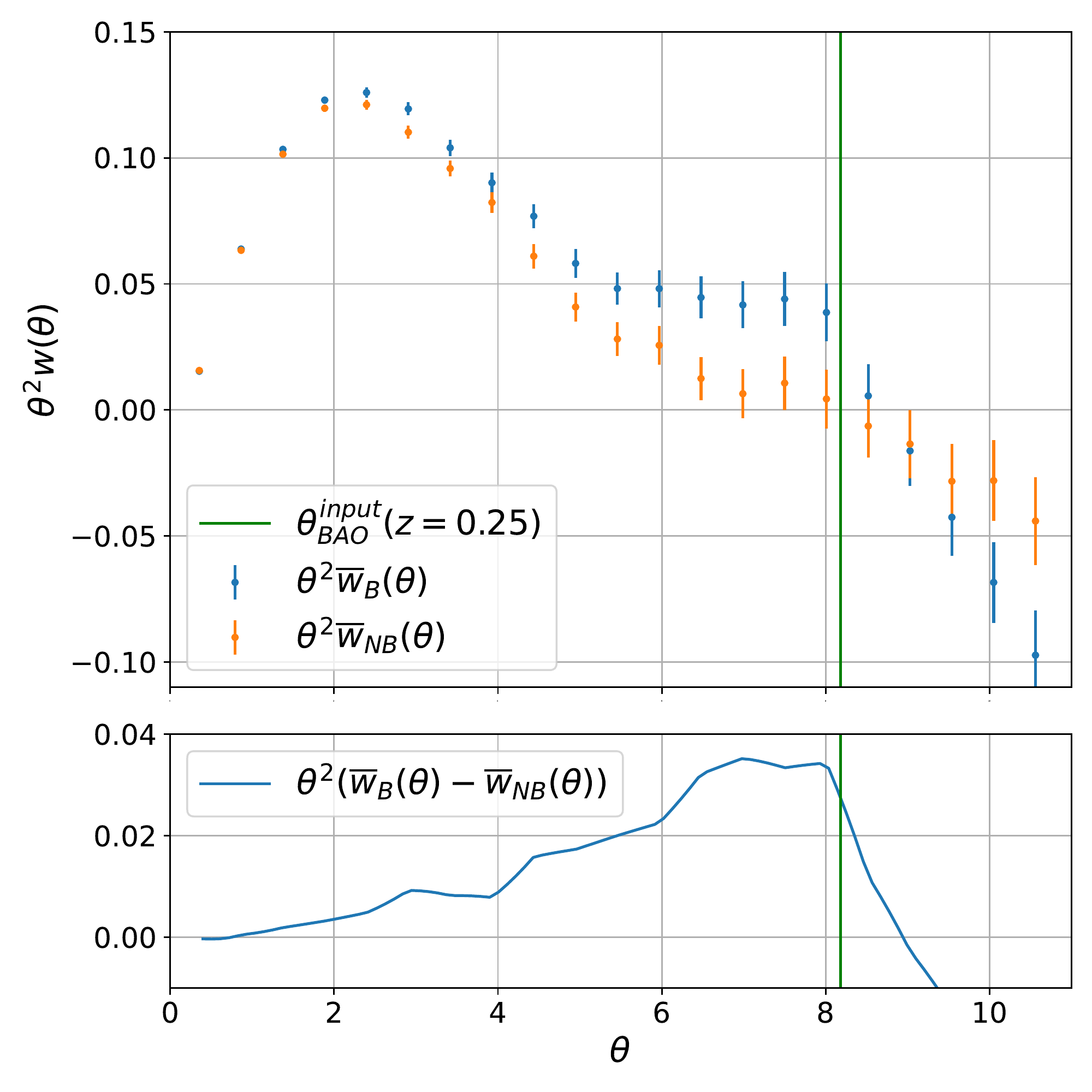}
\hfill
\includegraphics{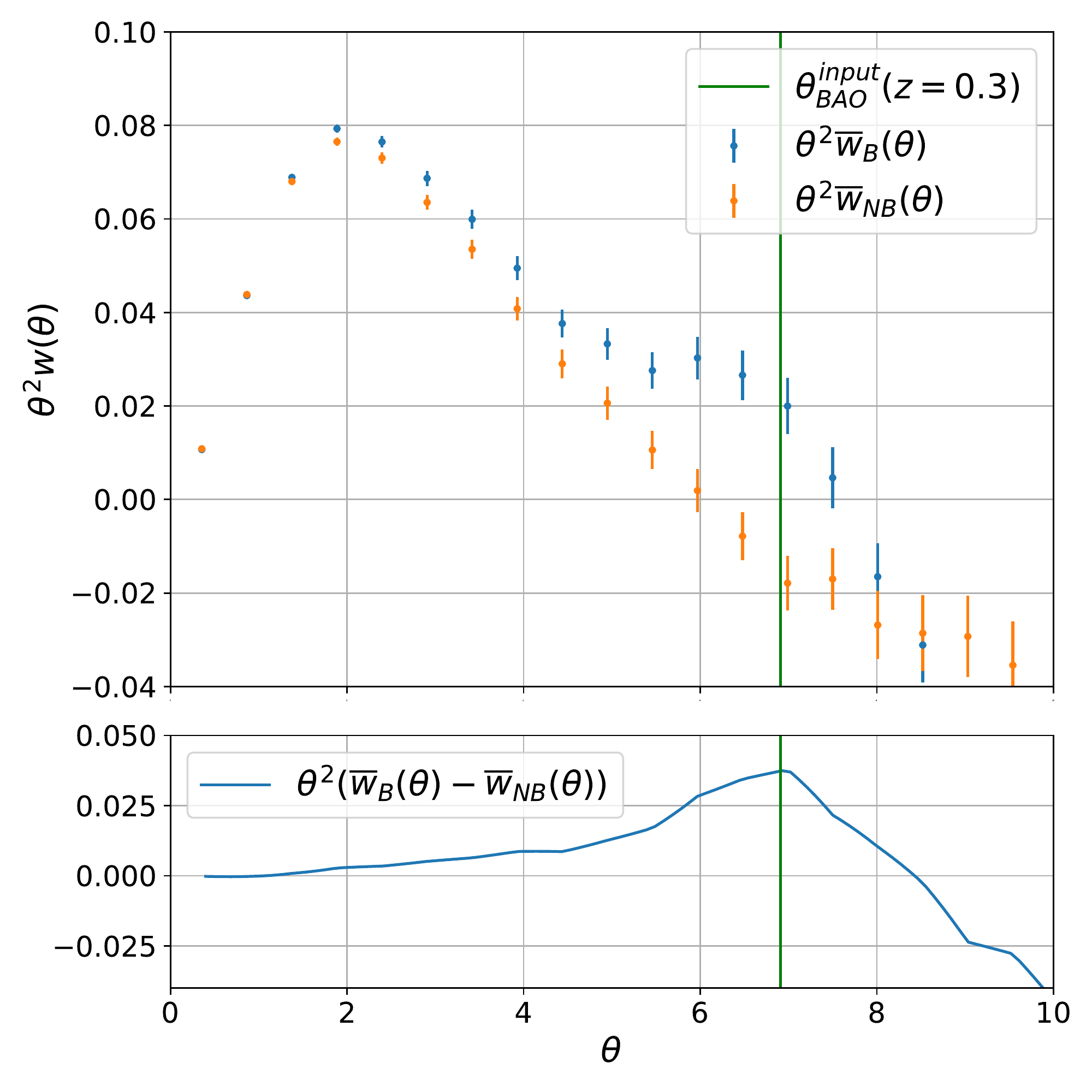}
\caption{The average of the 2PACF for all the BNS catalogues. $\overline{w}_{B}(\theta)$ represents average for all the catalogs which contain BAO peak and $\overline{w}_{NB}(\theta)$ is the same for the catalogs with no BAO peak. In the upper panels, we show the 2PACF for both set of catalogs. The solid vertical line shows the input angular BAO scale at the given redshift. In the lower panel, we show the difference in the average correlation functions obtained from both set of catalogs. We show the quantity $\theta^2\overline{w}(\theta)$ on the vertical axis for better visualization of the BAO peak. The three figures are shown for redshifts $z=0.2$ (left panel), $0.25$ (middle panel) and $0.3$ (right panel).}
\label{fig:average_all_catalogues}
\end{figure*}

\begin{itemize}
    \item For a given redshift, choose a shell of thickness $\sim$ 150 $h^{-1}$ Mpc in comoving volume. To avoid the autocorrelation of points from the same posterior, randomly select one point from each posterior which lies within the chosen shell. {This is the optimal choice, as a larger shell will wash away the BAO peak and a smaller shell will not have enough events to estimate the 2PACF with enough precision.}
    \item Use the selected points to calculate 2PACF using the Landy-Szalay estimator \textit{ie}.,
    \begin{equation}
        w_i(\theta) = \dfrac{DD_i(\theta) - 2 DR_i(\theta) + RR_i (\theta)}{RR_i(\theta)} \text{ ,}
    \end{equation}
    where $DD_i(\theta)$ is the number of pairs of data points in the bin separated by angle $\theta$,  $RR_i(\theta)$ is the number of point-pairs in an equal-sized random catalog separated by $\theta$, and $DR_i(\theta)$ is the number of data-random pairs separated by $\theta$. We use the publicly available code \textsc{Corrfunc} \citep{Sinha:2019bfk, Sinha:2019reo} to calculate the correlation function. To minimize the projection effects in the shell, we divide shell of ~150 $h^{-1}$ Mpc into smaller sub-shell of 60 $h^{-1}$ Mpc with sliding window of 30 $h^{-1}$ Mpc and take the average.
    \item Repeat the above procedure for different realizations of posterior field (by randomly selecting a point from each posterior) and estimate the average 2PACF $w(\theta) = \frac{1}{n}\sum_{i=1}^n w_i(\theta)$. For this study, we chose $n=100$ for each sub-shell of 60 $h^{-1}$ Mpc.
    \item Once we have recover $w(\theta)$, we model the 2PACF following \citep{Sanchez:2010zg} as,
    \begin{equation}
        w(\theta) = A + B\theta^\nu + C \exp[-\frac{\qty(\theta-\theta_{FIT})^2}{2\sigma_{FIT}^2}] \label{eqn:BAO_fit}
    \end{equation}
    This model has six parameters to fit the data: \{A, B, C, $\nu$, $\theta_{FIT}$, $\sigma_{FIT}$\}. {The first two terms in the model gives the power law to fit the broad shape of the correlation function and the last term models the BAO peak as a Gaussian with location of peak as $\theta_{FIT}$ and width of the peak as $\sigma_{FIT}$ along with amplitude C.} To fit a model without BAO peak, we drop the last term in \ref{eqn:BAO_fit} and fit for remaining three parameters.
    \item To account for systematic and statistical errors, we generate 50 galaxy catalogs for each redshift corresponding to different seed for underlying density field. We then take 20 realizations from each galaxy catalog to account for statistical fluctuation in choosing the set of host galaxies. In this way we account for errors due to cosmic variance, errors arising from the sampling bias due to the selection of the host galaxies, and errors due to localization posteriors.
\end{itemize}

We do not correct the recovered 2PACF $w(\theta)$ for i) smearing effects due to localization errors, and ii) projection effects in a shell. This is justified because we do not track the exact shape of 2PACF. Rather, we are interested in the location of BAO peak in the 2PACF. As long as these effects do not destroy the BAO peak in the correlation function, we should be able to recover it. The recovery of BAO peak with BNS merger events is also a statistical effect: one can confuse a statistical bump in correlation function with BAO peak. In order to be confidently recover the BAO peak, one must consider recovering BAO peak in various redshift bins.

\subsection{Results}
In figure \ref{fig:w_theta_recovered}, we show the recovery of the BAO peak in one realization of BNS merger events at the redshift $z=0.3$. We estimate the input angular BAO scale at given redshift $z$ using the relation $\theta_{BAO} = {r_s}/({(1+z)D_A(z)})$ where $r_s$ is the BAO scale in terms of comoving distance and $D_A(z)$ is the angular diameter distance to given redshift. For this BNS catalog, we fit the models with and without a BAO peak using \textsc{dynesty} \citep{speagle:2019} software package. We estimate the Bayesian evidence $\mathcal{Z}$ for both the models and compare them. The model with higher value ($>2.5$) of $\mathcal{Z}$ is statistically preferred \citep{jeffreys1998theory}. The difference in log evidence between the two models turns out to be $\ln{(\mathcal{Z}_1/\mathcal{Z}_2)} = 2.59$ indicating that the model with a BAO peak is favoured compared to model without a BAO peak.

In figure \ref{fig:bao_recovery} we show the the recovery of the BAO peak at different redshift bins. We estimate the Bayesian evidence for both models in these redshift bins. The $\ln{(\mathcal{Z}_1/\mathcal{Z}_2)}$ for these redshift bins is given by $23.29 (z=0.2)$, $16.73 (z=0.25)$, and $2.59 (z=0.3)$ again indicating that model with a BAO peak is favoured. We show that with 7-10 years of observation, enough BNS merger events can be accumulated to recover the BAO peak within statistical errors.

{For these simulations, the significant budget in errors arise from the sampling bias and due to cosmic variance. Scatter due to posterior samples contribute the least in the error budget. Hovewer, this would change if we go to higher redshift where localization errors become dominant due to large scatter in localization posteriors. The errors due to sampling bias will decrease when number of detections are increased, for example, due to higher merger rates.}

To check the robustness of the method, we also generate $\sim$ 1000 BNS catalogs from the corresponding galaxy catalogs which do not contain BAO peak. We use BBKS transfer function \citep{1986ApJ...304...15B} to calculate input correlation function to generate such catalogs. We then estimate the average 2PACF $w(\theta)$ across all catalogs in respective catagories (with BAO peak and without BAO peak). In figure \ref{fig:average_all_catalogues}, we show that the average of 2PACF estimated from all the BNS mereger catalogs in two catagories: i) ones containing a BAO peak and ii) ones that do not a BAO peak. It can be seen that, statistically, we recover the BAO peak at the injected value. In these simulations, we find that the redshift window between $z \in  \left[0.2,0.3\right]$ is best suited for our study because we get a large number of BNS events with desired localization accuracy. Beyond $z>0.3$, although we do get enough number of BNS events localized with a degree square, the localization errors along the radial direction start to dominate and become large enough to destroy the angular correlations as well. Since the BAO feature is very weak and is hard to detect, to account for statistical fluctuations, it is preferable to recover the BAO feature in 2PACF in a sliding window of a given shell thickness in the ideal redshift range described above.

These measurements in gravitational wave catalogs, apart from being independent probes of the BAO scale, provide the opportunity to constrain cosmological parameters by using the BAO scale as a standard ruler. At the low redshifts of interest to this study, $r_s$ is a direct measure of the Hubble parameter $H_0$. Using the $r_s$ and  $\Omega_m h^2$ (where $h = H_0/100$) derived from CMB experiments in conjuction with the measurements from GW data, one can measure the value of $\Omega_m$ \citep{Eisenstein:1998tu}. Alternatively, the measurements of the BAO scale $\theta_{BAO}$ at different redshifts can be used to measure $r_s$ \citep{Carvalho:2015ica}.

Although we use a GW detector network consisting only of 3G detectors, it is also possible that many current ground based detectors will still be in operation (with future upgrades). This scenario will only improve the localization of sources and hence a hybrid network consisting of 3G detectors such as CE, ET and 2G detectors such as LIGO, Virgo, KAGRA will greatly improve the localization of the GW sources. 

In this study, we assumed that the network of detectors will have same sensitivity for all sky positions. Depending on the given network configuration and antenna pattern, we might get varying sensitivity for different parts of sky. Although for 3G detectors, we expect this effect to be small but one natural extension of this work is to include such effects. In future, we intend to extend this work to include smearing effects due to posteriors \citep{Vijaykumar:2020pzn}, projection effects due to shell thickness \citep{Limber:1954zz}, and more current generation detectors along with 3G detector networks. The conclusions of the current work also rely on the range of estimated merger rates of BNS events \citep{LIGOScientific:2020ibl}. Future increase (decrease) in the estimation of merger rates would mean less (more) time will be required to accumulate enough BNS merger events to probe BAO. 

\section{Summary}\label{sec:summary}
We explore the possibility of detecting BAO scale using GW merger events in the 3G detector network. Probing the details of large scale structures (such as BAO scale) with GW observations is a challenging task because of poor localization of the GW sources and low number density of detected events. We find that with 3G detector network consisting of two CE (USA and Australia) and one ET (Europe), we can accumulate a large number ($\mathcal{O}(10000)$) of very well localized (within 1 square degree) BNS events upto the redshift ($z<0.3$) in 7-10 years of observing time. This opens up the possibility to probe the BAO scale solely by GW observations and hence provide an independent probe to BAO. With the 3G detector network considered in this study, we find that the redshift range of $z \in [0.2,0.3]$ is best suited for recovery of BAO peak assuming that the GW merger population does track the galaxy distribution.   We showed this through simulations at three different redshifts---0.2, 0.25, 0.3---by considering a small shell of thickness $~150 h^{-1}$ Mpc is centered around each redshift. In reality, when we accumulate enough BNS events in this redshift range, we can divide it into many smaller redshift bins to estimate the 2PACF and infer the presence of the BAO peak in each bin. The new probe for BAO will not only complement the observations from other surveys, it may provide opportunity to peek into the distribution of BNS with relation to galaxies and provide independent constraints on cosmological parameters.  This study broadens the horizon of science goals which can be achieved by 3G detectors and emphasizes the need for 3G detector network for the future. 

{We made a few simplifying assumptions for the purpose of this work. As a proof-of-concept, we only considered BNSs and their localization at low redshifts. We reiterate that this choice is based purely on the measured relative intrinsic merger rates of BBHs and BNSs, and their localization volumes. If the numbers and localization volumes of BBHs are comparable to those considered in this study at some redshifts, the methods described will translate trivially.}

{We also assumed that all galaxies in our mock catalog would host BNS events with equal probability; however, more massive/luminous galaxies are expected to be preferred hosts for BNS events. Hence, ideally, one should have weighted the galaxies by their mass while populating them with BNSs. We expect the mass weighting to affect the measured large scale-bias of GW events since the bias is known to strongly depend on of galaxy luminosity \citep{SDSS:2004oes}. However, we expect the errors introduced on the position and shape of the BAO peak (less than a percent)  \citep{Smith:2006ne} to be subdominant to the measurement errors in the 2PACF.}

{Lastly, we restricted ourselves to using the 2PACF in order to measure clustering. An equivalent analysis could also be performed using angular power spectra \citep{Peebles:1973}, as is done for other cosmological probes. We plan to investigate this thoroughly, along with the effects mentioned in the preceding paragraphs, in future work.}

\acknowledgments
We acknowledge the Max Planck Gesellschaft, and thank the computing team from AEI Hannover for their significant technical support. Authors thank Ajith Parameswaran, Tirthankar Roy Choudhury, Bruce Allen, and Badri Krishnan for useful discussions and valuable comments, and the anonymous referee for a careful reading of the draft and many useful suggestions. AV would also like to thank members of the Astrophysical Relativity group at ICTS for feedback. SK would like to thank Xisco Jim\'enez Forteza for useful comments. AV's research is supported by the Department of Atomic Energy, Government of India, under Project No. RTI4001.

\bibliography{references}

\end{CJK*}
\end{document}